\documentclass[12pt,aps,prd,preprint,tightenlines,superscriptaddress,
   showpacs,nofootinbib]{revtex4-1}
\usepackage{graphicx}

\usepackage{amsmath}
\usepackage{array}

\newcommand{\PRE}[1]{{#1}} % Use if preprint style

\newcommand{\nbox}{{\,\lower0.9pt\vbox{\hrule \hbox{\vrule height 0.2 cm
\hskip 0.2 cm \vrule height 0.2 cm}\hrule}\,}}

\newcommand{\ifb}{\text{fb}^{-1}}

\newcommand{\mev}{\text{MeV}}
\newcommand{\gev}{\text{GeV}}
\newcommand{\tev}{\text{TeV}}

\newcommand{\eg}{{\em e.g.}}

\newcommand{\be}{\begin{equation}}
\newcommand{\ee}{\end{equation}}
\newcommand{\bea}{\begin{eqnarray}}
\newcommand{\eea}{\end{eqnarray}}
\newcommand{\baln}{\begin{align}}
\newcommand{\ealn}{\end{align}}
\newcommand{\lsim}{\lower.7ex\hbox{$\;\stackrel{\textstyle<}{\sim}\;$}}
\newcommand{\gsim}{\lower.7ex\hbox{$\;\stackrel{\textstyle>}{\sim}\;$}}

\DeclareMathOperator{\Tr}{Tr}

\usepackage{graphicx}
\begin{document}

\preprint{UH-511-1200-2012}
\preprint{UCI-TR-2012-14}

\title{
\PRE{\vspace*{1.3in}}
Spin Determination for Fermiophobic Bosons
\PRE{\vspace*{0.3in}}
}

\author{Jason Kumar}
\affiliation{Department of Physics and Astronomy, University of
Hawai'i, Honolulu, HI 96822, USA
\PRE{\vspace*{.1in}}
}

\author{Arvind Rajaraman%
\PRE{\vspace*{.4in}}
}
\affiliation{Department of Physics and Astronomy, University of California, Irvine, CA  92697
\PRE{\vspace*{.5in}}
}

\author{David Yaylali}
\affiliation{Department of Physics and Astronomy, University of
Hawai'i, Honolulu, HI 96822, USA
\PRE{\vspace*{.1in}}
}

%\date{June 2012}

\begin{abstract}
\PRE{\vspace*{.3in}}
We discuss the prospects for production and detection of fermiophobic bosons 
(exotic bosons decaying only to standard model gauge bosons) at the LHC, and 
describe simple methods for determining spin.
We consider two complementary approaches to spin determination: the search for 
decays in the diphoton channel, and the comparison
of events with no extra spectator jets to those with one extra jet.
We show that these approaches can together allow the fermiophobic boson's spin to be determined over a
wide range of parameter space.  We study
both even and odd parity states.

\end{abstract}

\maketitle
%\newpage

\section{Introduction} \label{sec:intro}

In this paper we will study the phenomenology of massive fermiophobic bosons (denoted throughout this paper as $X$).
Fermiophobic bosons dominantly couple to standard model gauge-bosons.
We consider possible strategies for distinguishing a spin-0 fermiophobic resonance from a spin-1
resonance (we consider both even and odd parity).  We will find that there are two effective complementary strategies:
a search for diphoton events, and
a comparison of events with no spectator jets to those with one spectator jet.

Fermiophobic bosons have been studied in a variety of
contexts~\cite{Holdom:1985ag,delAguila:1995rb,Dienes:1996zr,Anastasopoulos:2006cz,Kumar:2006gm,Feldman:2006wb,Chang:2006fp,Chang:2007ki,
Feldman:2007wj,Kumar:2007zza,Antoniadis:2009ze,Fermiophobic:2011,Bach:2011jy}, particularly in the context of a fermiophobic Higgs
boson~\cite{FermiophobicHiggs}.  Fermiophobic bosons can
arise naturally in models where a hidden sector couples dominantly to exotic heavy fermions or scalars.  If these 
matter fields are heavy enough,
then they cannot be directly produced at the LHC.  But if these fields are charged under standard model gauge groups,
then loops of heavy particles mediate an effective coupling to standard model gauge bosons.  One-loop diagrams can thus mediate
mixing between standard model gauge bosons and fermiophobic bosons, and search strategies for this effect have
been well-studied~\cite{Holdom:1985ag,delAguila:1995rb,Dienes:1996zr,Kumar:2006gm,Feldman:2006wb,Chang:2006fp,Chang:2007ki,Feldman:2007wj}.
An alternative strategy is to search for triple-boson couplings, where $X$
couples to two standard model gauge bosons.  This strategy can be especially useful in cases where $X$ does not mix
with standard model gauge bosons; examples include the case where $X$ is a scalar, or when the heavy fields mediating
the interaction are not charged under U(1)$_Y$.

One would expect that the most effective way to search for fermiophobic bosons with triple boson couplings would
be through the process $gg \rightarrow X \rightarrow VV$, where $V$ is an electroweak gauge boson.  This process
benefits from large production rate associated with a gluon initial state, and clean photon or lepton signals which
are possible with an electroweak boson final state.  This channel was studied in \cite{Fermiophobic:2011} for the case where
$X$ is a pseudovector.  A variety of experimental searches for fermiophobic bosons have been conducted~\cite{FermiophobicExpSearch}.

It is important to determine all quantum numbers of any exotic resonance discovered at the LHC.
The measurement of the spin of a resonance has been studied by many authors, particularly in
the context of resonances similar to the Higgs \cite{HiggsImpostors}.
A typical method of spin determination
is to look at the angular distribution of the decay products in the rest frame of the decaying particle.
We will consider two complementary approaches to spin determination which do not require the use of angular information, but
instead rely on effects associated with Landau-Yang theorem~\cite{Keung:2008ve,LandauYang} selection rules.

According to the Landau-Yang theorem, a massive vector boson cannot decay to two identical massless vector bosons.
Thus, if the decay $X \rightarrow \gamma \gamma$ is observed, it is clear that $X$ is not spin-1.
This is the reason why it is clear that the recent exciting discovery at the LHC of a boson with a mass of $\sim 125~\gev$~\cite{Higgs} cannot
be a spin-1 resonance.
The Landau-Yang theorem
also implies that a massive vector cannot be produced from an initial state of two on-shell gluons.  This implies
that the production of a spin-1 fermiophobic boson from an $Xgg$ vertex must be accompanied by the emission of a
spectator jet.
We will find that the these two strategies can be used together to distinguish between spin-0 and \mbox{spin-1} fermiophobic
bosons for a wide range of models.  The efficacy of each strategy depends primarily on the relative strength
of the coupling of $X$ to SU(2) and to U(1)$_Y$ gauge bosons.

In section II, we describe the coupling of fermiophobic bosons to standard model gauge bosons in terms
of higher-dimension effective operators.  In section III we describe the primary $X$ production and
decay processes.  In section IV we describe our method of simulating signal and background events, and
in section V we describe the sensitivity of the LHC, and its ability to distinguish the spin of
fermiophobic resonances.  We conclude in section VI with a discussion of our results.

\section{Effective theory of the fermiophobic boson}\label{sec:theory}

The theory we consider consists of a new massive boson $X$ which has effective couplings to standard model gauge bosons.  We
will assume $X$ is not charged under standard model gauge groups; the coupling can thus be written as an effective operator of
dimension 5 or 6 (depending on the spin of $X$) which couples $X$ to standard model field strengths.
We now list the possible (lowest dimensional) effective operators for the four spin/parity assignments of this boson.

For either a scalar or pseudoscalar, there is one lowest dimension (dimension 5) effective operator.
\begin{eqnarray}
{\cal O}_s &=& \frac{1}{\Lambda}X \Tr [F_{\mu \nu} F^{\mu \nu} ],
\\
{\cal O}_{ps} &=& \frac{1}{\Lambda} \epsilon_{\alpha \beta \gamma \delta}X \Tr [F^{\alpha \beta} F^{\gamma \delta}].
\end{eqnarray}

For a (pseudo)vector $X$, the lowest dimensional operator is of dimension 6.  For the vector, we find four possible operators:
\begin{align}
{\cal O}_v^1 &= {1 \over \Lambda^2} X^\mu \Tr [F_{\mu \alpha} \partial_\beta F^{\alpha \beta} ] \\
{\cal O}_v^2 &= {1 \over \Lambda^2} X^\mu \Tr [\partial_{\beta} F_{\mu \alpha}  F^{\alpha \beta} ] \\
{\cal O}_v^3 &= {1 \over \Lambda^2} \partial_\beta X^\mu \Tr [F_{\mu \alpha} F^{\alpha \beta} ] \\
{\cal O}_v^4 &= {1 \over \Lambda^2} X^\mu \Tr [F_{\alpha \beta} \partial_{\mu} F^{\alpha \beta} ],
\end{align}
where we have written each operator only to quadratic order in standard model gauge field strengths.  For a fully
gauge-invariant operator, the partial derivative should be replaced by a covariant derivative.
The first three operators are related by an integration by parts.
\[
X^\mu \Tr [F_{\mu \alpha} \partial_\beta F^{\alpha \beta}] = \partial_{\beta} (X^\mu \Tr [F_{\mu \alpha}
F^{\alpha \beta}]) - (X^\mu \Tr [\partial_{\beta} F_{\mu \alpha}  F^{\alpha \beta}] + \partial_\beta X^\mu
\Tr [F_{\mu \alpha} F^{\alpha \beta}]).
\]
Dropping the surface term, we find
\[
{\cal O}_v^1 \sim -({\cal O}_v^2+{\cal O}_v^3).
\]
We will in what follows assume $X$ to be on-shell.  In this case, we find that the operator ${\cal O}_v^2$ does not contribute.
The vertex function for this operator is
\begin{eqnarray}
\Gamma^{\mu \nu \rho}_{v(2)}(k_X,k_1,k_2) &=& \frac{1}{\Lambda^2}(k_1^\mu + k_2^\mu)(k_1^\rho k_2^\nu-g^{\rho \nu}k_1 \cdot k_2)
\nonumber\\
&=& \frac{1}{\Lambda^2} k_X^\mu (k_1^\rho k_2^\nu-g^{\rho \nu}k_1 \cdot k_2).
\end{eqnarray}
Contracting this vertex function with the polarization vector for the $X$ thus gives zero.  The vertex function for ${\cal O}_v^4$ is identical, and
so also does not contribute.  Thus the only operators which contribute to the $X$ coupling are ${\cal O}_v^1$ and ${\cal O}_v^3$, which are
equivalent by an integration by parts.  We thus drop superscript labels and write
\begin{align}
{\cal O}_v=\frac{1}{\Lambda^2} X^\mu \Tr [F_{\mu \alpha} \partial_\beta F^{\alpha \beta} ].
\end{align}

Finally, for the pseudovector $X$, the only non-vanishing operator is~\cite{Fermiophobic:2011}
\begin{align}
{\cal O}_{pv}=\frac{1}{\Lambda ^2}X^\mu \epsilon_{\mu \nu \rho \sigma} \Tr [F^{\nu \rho} \partial_\beta F^{\beta \sigma} ] .
\end{align}
For non-Abelian fields $F_{\alpha \beta}^a$, covariant derivatives should be used to act on the field
strengths.  The covariant derivatives will give rise to higher point vertices, allowing decays to 3 or more standard model bosons.\footnote{Extra
terms arising from the covariant derivative will not affect two-body decay widths.  They do, however, affect two-body branching fractions through
their contribution to three-body widths, and so we use the full covariant derivatives in all of our couplings for event generation.}
For coupling to gluons, the higher point interactions have a sizable contribution
to the production channel.

We will assume a higher-dimensional effective coupling to all gauge groups of the standard model,
though for generality we will allow the couplings to be different.
We will thus characterize the couplings by an energy scale $\Lambda$, and encode the relative strength of the coupling to SU(2)$_L$ and
U(1)$_Y$ in relation to SU(3)$_{\rm QCD}$ by $C_1$ and $C_2$.  The effective operators are thus given as
\begin{align}
{\cal O}_s &= \frac{1}{\Lambda_{s}}XG_{\mu \nu}^{a}G_{a}^{\mu \nu}+\frac{C_1}{\Lambda_{s}}XW_{\mu \nu}^{i}W_{i}^{\mu \nu}
+\frac{C_2}{\Lambda_{s}}XB_{\mu \nu}B^{\mu \nu} \label{eq:OPs}\\
{\cal O}_{ps} &= \frac{1}{\Lambda_{ps}}XG^{a}_{\mu \nu}\widetilde{G}^{\mu \nu}_a
+\frac{C_1}{\Lambda_{ps}}XW^{i}_{\mu \nu}\widetilde{W}^{\mu \nu}_i+\frac{C_2}{\Lambda_{ps}}XB_{\mu \nu} \widetilde{B}^{\mu \nu} \label{eq:OPps} \\
{\cal O}_{v} &= \frac{1}{\Lambda_{v}^2}X^{\mu}G_{\mu \alpha}^{a}D_{\beta}G^{a\alpha \beta}
+\frac{C_1}{\Lambda_{v}^2}X^{\mu}W_{\mu \alpha}^{i}D_{\beta}W^{i\alpha \beta}
+\frac{C_2}{\Lambda_{v}^2}X^{\mu}B_{\mu \alpha}D_{\beta}B^{\alpha \beta} \label{eq:OPv}\\
{\cal O}_{pv} &= \frac{1}{\Lambda_{pv}^2}X^{\mu}\widetilde{G}_{\mu \alpha}^{a}D_{\beta}G^{a\alpha \beta}
+\frac{C_1}{\Lambda_{pv}^2}X^{\mu}\widetilde{W}_{\mu \alpha}^{i}D_{\beta}W^{i\alpha \beta}
+\frac{C_2}{\Lambda_{pv}^2}X^{\mu}\widetilde{B}_{\mu \alpha}D_{\beta}B^{\alpha \beta} \label{eq:OPpv}
\end{align}
with $\widetilde{F}_{\mu \nu}=\tfrac{1}{2}\epsilon_{\mu \nu \sigma \tau}F^{\sigma \tau}$.  Since field strengths are parity even
objects and $\epsilon_{\mu \nu \sigma \tau}$ is parity odd, we see that the $X$ in Eq.~(\ref{eq:OPps}) and Eq.~(\ref{eq:OPpv})
is a pseudoscalar and pseudovector, respectively.

\section{Production and Decay Channels} \label{sec:ProdAndDecay}

\begin{figure}[htp]
\centering
\includegraphics[scale=0.70]{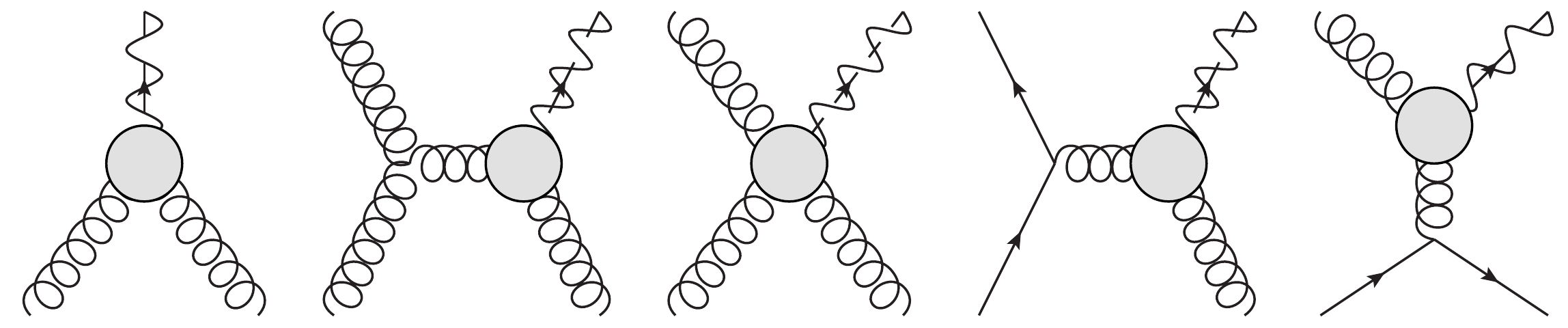}
\caption{Partonic production channels for the X}
\label{fig:production}
\end{figure}

Production at the LHC will be primarily through the channels $gg$, $gq$, $g \bar{q}$, and $q \bar{q}$.  Vector boson fusion
is also a viable production channel for the $X$~\cite{Kumar:2007zza}, but will only become important when coupling to SU(3) is small or zero.
 Thus we will not consider VBF as a production channel in this analysis, though we note that inclusion of this production
 channel will only add to the 2-jet cross section.  We see (Fig.~\ref{fig:production}) that all but the $gg$ partonic production
 channels require the presence of a spectator jet.
If $X$ is spin-1, the dominant $gg \rightarrow X$ production channel is forbidden for on-shell $X$; thus spin-1 $X$ is
{\it always} produced with at least one accompanying jet.  Since this dominant production channel is available
to the spin-0 $X$, we should in principle be able to determine spin by simply comparing the zero-jet and one-jet
cross sections for any observed fermiophobic resonance.  We will return to this possibility in a later section.
Interestingly, it turns out that the matrix element for the $gg \rightarrow Xg$ hard process also turns out to
be zero if $X$ is spin-1, so production occurs only through quark-gluon and quark-antiquark scattering.

We list the production cross sections for the different spin/parity assignments in Table~\ref{ProdXsec}, as calculated in MadGraph5.
Cross sections for production with zero and one jet are listed separately.  We see, as expected, that the cross section for spin-1 $X$
vanishes when there is no accompanying jet.  We also see that production for spin-1 $X$ is parity dependent at low masses, while for
spin-0 $X$ there is no noticeable dependence.
\begin{table}
\caption{Tree-level production cross sections (in fb) for $X+\mbox{0-jet}$ and $X+\mbox{1-jet}$,  as calculated by 
MadGraph5 for $C_1 =10$, $C_2 =0$, and either
$\Lambda =50~\tev$ (spin-0) or $\Lambda =\sqrt{10}~\tev$ (spin-1).  Cross sections for parity even and parity odd are 
identical for a spin-0 $X$.  Jets are required to
have $p_{T} \geq 50~\gev.$
}
\label{ProdXsec}
\begin{tabular*}{0.75\textwidth}{@{\extracolsep{\fill}} l | c c c c}
	\hline \hline
	$m_X$	&	$500~\gev$	&	$1000~\gev$		&	$1500~\gev$		&	$2000~\gev$  \\
	\hline
	$\sqrt{s}=8~\tev$ \\
	$\sigma(p p \rightarrow X_{(p)s})$  	& $4.70\times10^{2}$ & $3.38\times10^{1}$ & $4.31$ & $6.82\times 10^{-1}$ \\
	$\sigma(p p \rightarrow X_{(p)v})$  	& 0 & 0 & 0 & 0	\\
	$\sigma(p p \rightarrow X_{(p)s}+j)$	& $2.71\times10^{2}$ & $2.69\times10^{1}$ & $4.03$ & $7.04\times10^{-1}$\\
	$\sigma(p p \rightarrow X_{v}+j)$  	& $7.26\times10^{1}$ & $8.17$ & $1.36$ & $2.51\times10^{-1}$ \\
	$\sigma(p p \rightarrow X_{pv}+j)$	& $2.51\times10^{1}$ & $8.17$ & $1.36$ & $2.51\times10^{-1}$ \\
	\hline
	$\sqrt{s}=14~\tev$ \\	
	$\sigma(p p \rightarrow X_{(p)s})$ 	& $2.25 \times 10^{3}$  & $2.76 \times 10^{2}$ & $5.91\times 10^{1}$   & $1.62\times 10^{1}$ 	\\
	$\sigma(p p \rightarrow X_{(p)v})$	& 0 & 0 & 0 & 0	\\
	$\sigma(p p \rightarrow X_{(p)s}+j)$	& $1.53 \times 10^{3}$ & $2.53 \times 10^{2}$ & $6.40 \times 10^{1}$ & $1.94 \times 10^{1}$ \\
	$\sigma(p p \rightarrow X_{v}+j)$ 	& $7.48\times10^{2}$ & $1.32\times10^{2}$ & $3.72\times10^{1}$ & $1.24\times10^{1}$ \\
	$\sigma(p p \rightarrow X_{pv}+j)$	 & $1.24\times10^{3}$ & $1.32\times10^{2}$ & $3.72\times10^{1}$ & $1.24\times10^{1}$ \\
	\hline \hline
\end{tabular*}
\end{table}

\subsection{Decay Channels}\label{sec:Decay}

Since the $X$ is in general coupled to all gauge groups of the standard model, there is some non-zero branching
fraction to both gluons and electroweak gauge bosons.  If $X$ decays via its coupling to gluons, the final state signal
would lie in the multi-jet events; two or more jets for a spin-0 $X$ and 4 or more jets for a spin-1 $X$ (the matrix
elements for $X\rightarrow gg,ggg$ vanish).  As this is a difficult experimental analysis, we focus instead on
electroweak decays, which lead to much cleaner signals at the
LHC.  Furthermore, since we want to study an on-shell resonance, it is necessary that we can fully reconstruct the $X$
from its decay products, and so we do not look in channels that contain neutrinos in the final state.
We will look in the channels $X \rightarrow Z Z \rightarrow 4l$
(the so-called {\it golden channel}), $X \rightarrow Z \gamma \rightarrow 2l+\gamma$, and
$X \rightarrow \gamma \gamma$ (the diphoton channel is forbidden for if $X$ is spin-1).

Scalar decay widths are given by
\begin{align}
{\Gamma_s(WW) \over m_X} &= \frac{C_1^2}{2 \pi} {m_X^2 \over \Lambda^2}
\sqrt{1-\frac{4m_W^2}{m_X^2}}\left(1-\frac{4m_W^2}{m_X^2}+6  \frac{m_W^4}{m_X^4}\right) \label{eq:swid} \\
{\Gamma_s(ZZ) \over m_X} &= \frac{(C_1\cos^2{\theta_{w}}+C_2\sin^2{\theta_{w}})^2}{4 \pi}
{m_X^2 \over \Lambda^2} \sqrt{1-\frac{4m_Z^2}{m_X^2}}
\left(1-\frac{4m_Z^2}{m_X^2}+6  \frac{m_Z^4}{m_X^4}\right) \\
{\Gamma_s(Z \gamma) \over m_X} &= \frac{(C_1-C_2)^2\cos^2{\theta_{w}}\sin^2{\theta_{w}}}{2 \pi}
{m_X^2 \over \Lambda^2} \left(1-\frac{m_Z^2}{m_X^2}\right)^3 \\
{\Gamma_s(\gamma \gamma) \over m_X} &= \frac{(C_1\sin^2{\theta_{w}}+C_2\cos^2{\theta_{w}})^2}{4 \pi }
{m_X^2 \over \Lambda^2} \label{eq:sWidAA}\\
{\Gamma_s(g g) \over m_X} &=  \frac{2 }{\pi } {m_X^2 \over \Lambda^2}
\end{align}
Note that the functional structures are identical to those of the standard model Higgs decay.

Pseudoscalar decay widths are given by
\begin{align}
{\Gamma_{ps}(WW) \over m_X} &= \frac{C_1^2 }{ 2 \pi }{m_X^2 \over \Lambda^2} \left(1-\frac{4m_W^2}{m_X^2}\right)^{3/2} \\
{\Gamma_{ps}(ZZ) \over m_X} &= \frac{(C_1\cos^2{\theta_{w}}+C_2\sin^2{\theta_{w}})^2 }{4 \pi } {m_X^2 \over \Lambda^2}
\left(1-\frac{4m_Z^2}{m_X^2}\right)^{3/2} \\
{\Gamma_{ps}(Z \gamma) \over m_X} &= \frac{(C_1-C_2)^2\cos^2{\theta_{w}}\sin^2{\theta_{w}}}{2 \pi}
{m_X^2 \over \Lambda^2} \left(1-\frac{m_Z^2}{m_X^2}\right)^3  \\
{\Gamma_{ps}(\gamma \gamma) \over m_X} &= \frac{ (C_1\sin^2{\theta_{w}}+C_2\cos^2{\theta_{w}})^2}{4 \pi }
{m_X^2 \over \Lambda^2} \\
{\Gamma_{ps}(g g) \over m_X} &=  \frac{2 }{\pi } {m_X^2 \over \Lambda^2}
\end{align}
It is interesting to note that the scalar and pseudoscalar partial widths become equal as $m_{Z,W} / m_X \rightarrow 0$.  This suggests that
all information concerning the parity of the $X$ is encoded in the longitudinal modes of the $W$ and $Z$.

Vector decay widths are given by,
\begin{align}
{\Gamma_{v}(WW) \over m_X} &= \frac{C_1^2 }{48 \pi }{m_X^4 \over \Lambda^4} {m_W^2 \over m_X^2}  \left(  1-\frac{4m_W^2}{m_X^2} \right)^{3/2} \\
{\Gamma_{v}(ZZ) \over m_X} &= \frac{(C_1\cos^2{\theta_w}+C_2\sin^2{\theta_w})^2 }{96 \pi }  {m_X^4 \over \Lambda^4}
{m_Z^2 \over m_X^2} \left(  1-\frac{4m_Z^2}{m_X^2} \right)^{3/2}  \\
{\Gamma_{v}(Z\gamma) \over m_X} &= \frac{(C_1-C_2)^2 \cos^2{\theta_w}\sin^2{\theta_w} }{96 \pi } {m_X^4 \over \Lambda^4} {m_Z^2 \over m_X^2}
\left(  1+\frac{m_Z^2}{m_X^2} \right)\left(  1-\frac{m_Z^2}{m_X^2} \right)^3
\end{align}

Lastly, pseudovector decay widths are given by,
\begin{align}
{\Gamma_{pv}(WW) \over m_X} &= \frac{C_1^2 }{48 \pi }
{m_X^4 \over \Lambda^4}{m_W^2 \over m_X^2}
\left(  1-\frac{4m_W^2}{m_X^2} \right)^{5/2} \\
{\Gamma_{pv}(ZZ) \over m_X} &= \frac{(C_1\cos^2{\theta_w}+C_2\sin^2{\theta_w})^2 }{96 \pi }
{m_X^4 \over \Lambda^4}{m_Z^2 \over m_X^2}
\left(  1-\frac{4m_Z^2}{m_X^2} \right)^{5/2}  \\
{\Gamma_{pv}(Z\gamma) \over m_X} &= \frac{(C_1-C_2)^2 \cos^2{\theta_w}\sin^2{\theta_w} }{96 \pi }
{m_X^4 \over \Lambda^4}{m_Z^2 \over m_X^2}
\left(  1+\frac{m_Z^2}{m_X^2} \right)
\left(  1-\frac{m_Z^2}{m_X^2} \right)^3 \label{eq:pvwid}
\end{align}
Again, we see that the parity-even and parity-odd widths become equal as $m_{Z,W} / m_X \rightarrow 0$.

\begin{figure}[h!]
\begin{tabular}{cc}
\includegraphics[width=2.2in]{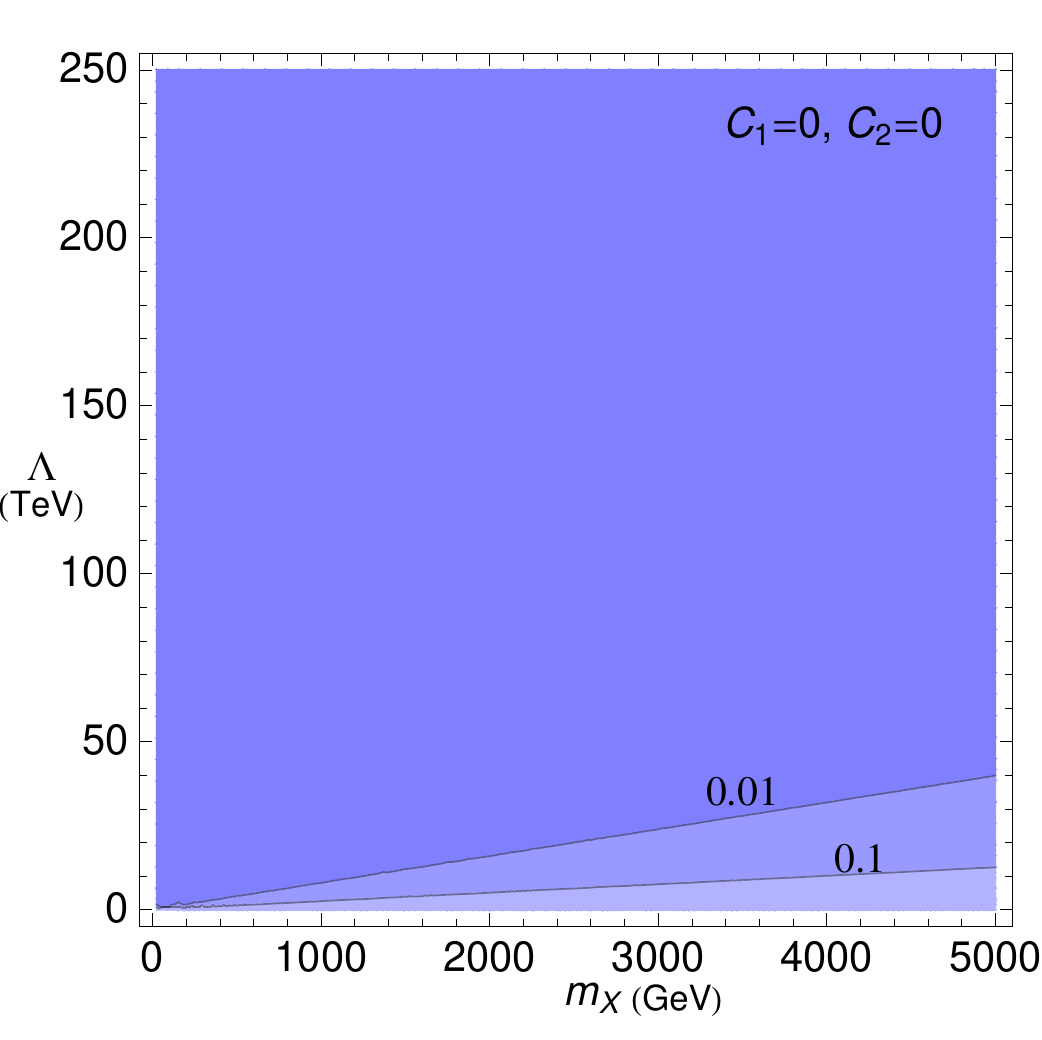} &
\includegraphics[width=2.2in]{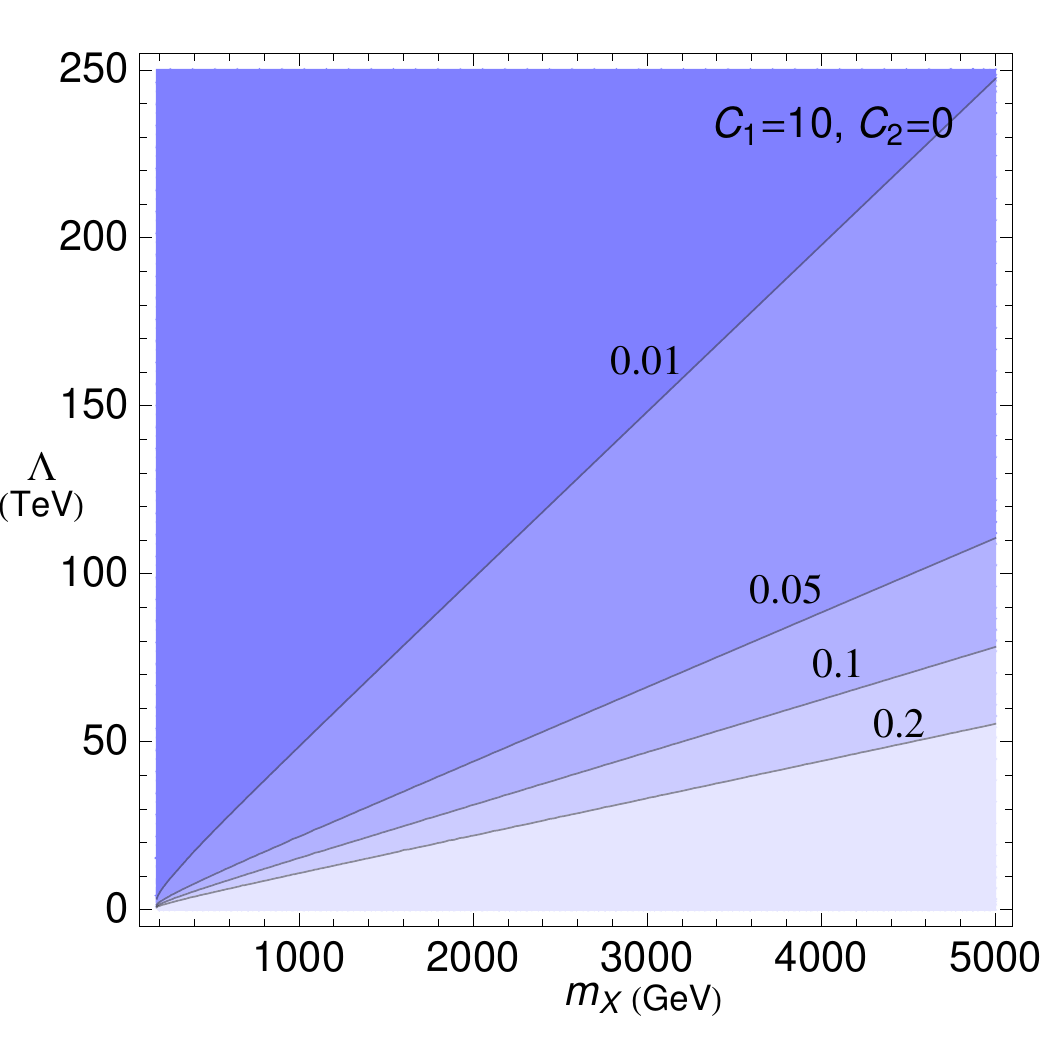} \\
\end{tabular}
\caption{Contours of $\Gamma_{tot} / m_{X}$ for spin-0 $X$, for coupling only to gluons ($C_1=C_2=0$) (left)
and for coupling to SU(2) with $C_1=10$, $C_2 =0$.
For a vast majority of the parameter space width is less than 10\% of the mass, and the narrow width approximation is valid.}
\label{fig:NarWidS}
\end{figure}

\begin{figure}[h!]
\includegraphics[width=2.2in]{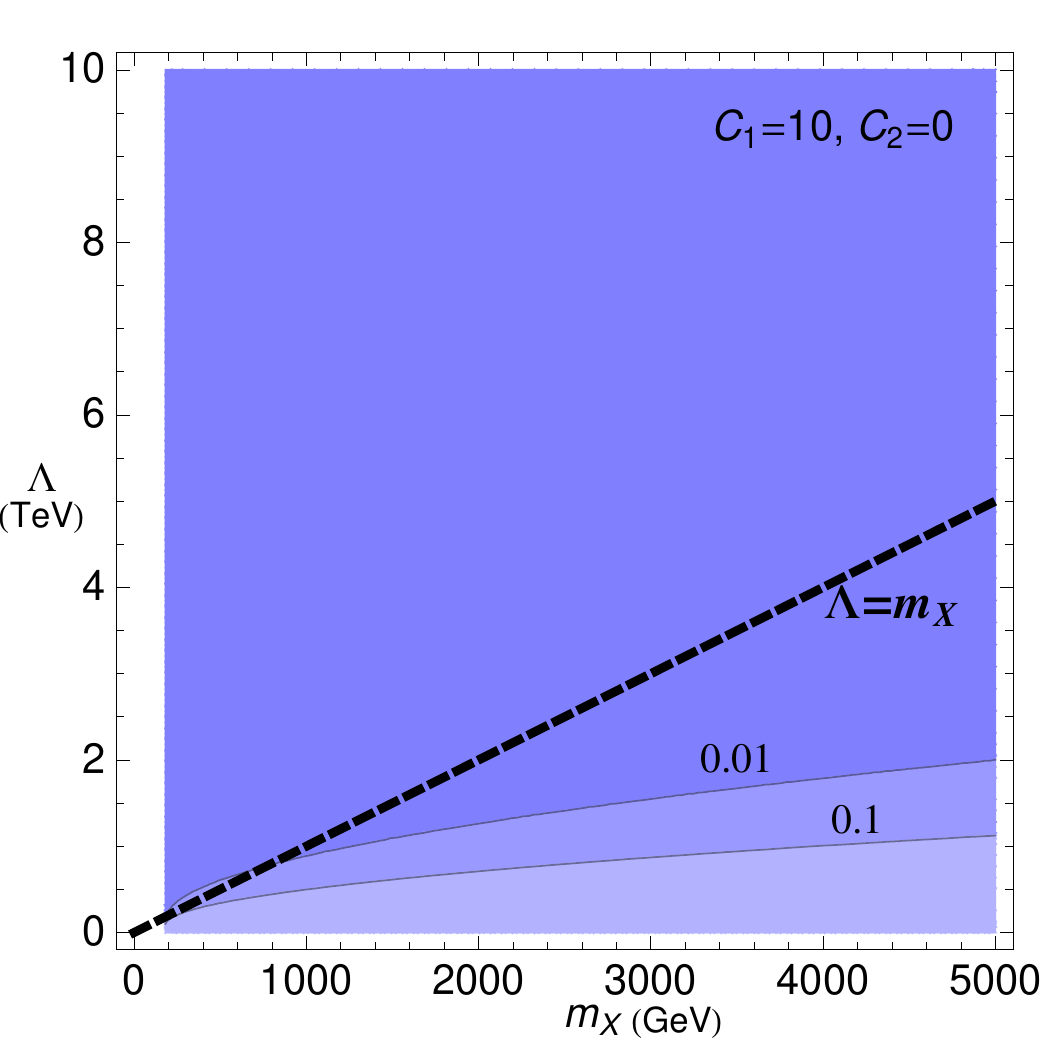}
\caption{Contours of $\Gamma_{tot} / m_{X}$ for spin-1 $X$.  Also shown is the line $\Lambda = m_{X}$,
at which point the effective field theory description is naively expected to break down.  For coupling to SU(2)
with $C_1=10$  we find $\Gamma_{tot} / m_X \ll 1$ over a vast majority of the parameter space, and the narrow width approximation holds.}
\label{fig:NarWidV}
\end{figure}

In Table~\ref{Wid}, we list the widths (in $\mev$) to each channel for $m_X = 500~\gev$, $C_1 =10$, $C_2 =0$, and either
$\Lambda =50~\tev$ (spin-0) or $\Lambda =\sqrt{10}~\tev$ (spin-1).
\begin{table}
\caption{Partial widths (in $\mev$), as calculated by MadGraph5, for $m_X = 500~\gev$, $C_1 =10$, $C_2 =0$, and either
$\Lambda =50~\tev$ (spin-0) or $\Lambda =\sqrt{10}~\tev$ (spin-1).
}
\label{Wid}
\begin{tabular*}{0.75\textwidth}{@{\extracolsep{\fill}} c | c c c c c c c}
	\hline \hline
	$~~V_1 V_2 ~~$ & $~~WW~~$ & $~~ZZ~~$ & $~~ Z \gamma ~~$ & $~~ \gamma \gamma ~~ $ & $gg ~~$ & {\it 3-body} & {\it total}\\
	\hline
	$\Gamma_{s}~$ & 680.2 & 190.1 & 128.8 & 21.73 & 31.83 & 89.63 & 1142\\
	$\Gamma_{ps}$ & 677.2 & 188.6 & 128.8 & 21.73 & 31.83 & 90.00 & 1138 \\
	$\Gamma_{v}~$ & 4.495 & 1.634 & 0.5762 & 0 & 0 & 0.3815 & 7.086 \\
	$\Gamma_{pv}$ & 4.037 & 1.416 & 0.5762 & 0 & 0 & 0.3516 & 6.381 \\
	\hline \hline
\end{tabular*}
\end{table}
We see that for all cases the narrow width approximation is valid.  To see that the narrow width approximation remains
valid throughout our parameter space, we plot the contours of $\Gamma_{tot} / m_{X}$ in $\Lambda$-$m_{X}$ space for both
spin-0 (Fig.~\ref{fig:NarWidS}) and spin-1 (Fig.~\ref{fig:NarWidV}).  We neglect any contributions from three-body decays.
We see that for spin-0, $\Gamma_{tot} / m_{X} < 10\%$ for a majority of the parameter space, though at large values
of $m_X / \Lambda$ the narrow width approximation may break down.   In the case of a spin-1 $X$, the narrow width approximation
holds over a much wider region of parameter space.

\begin{figure}[h!]
\begin{tabular}{cc}
\includegraphics[width=3in]{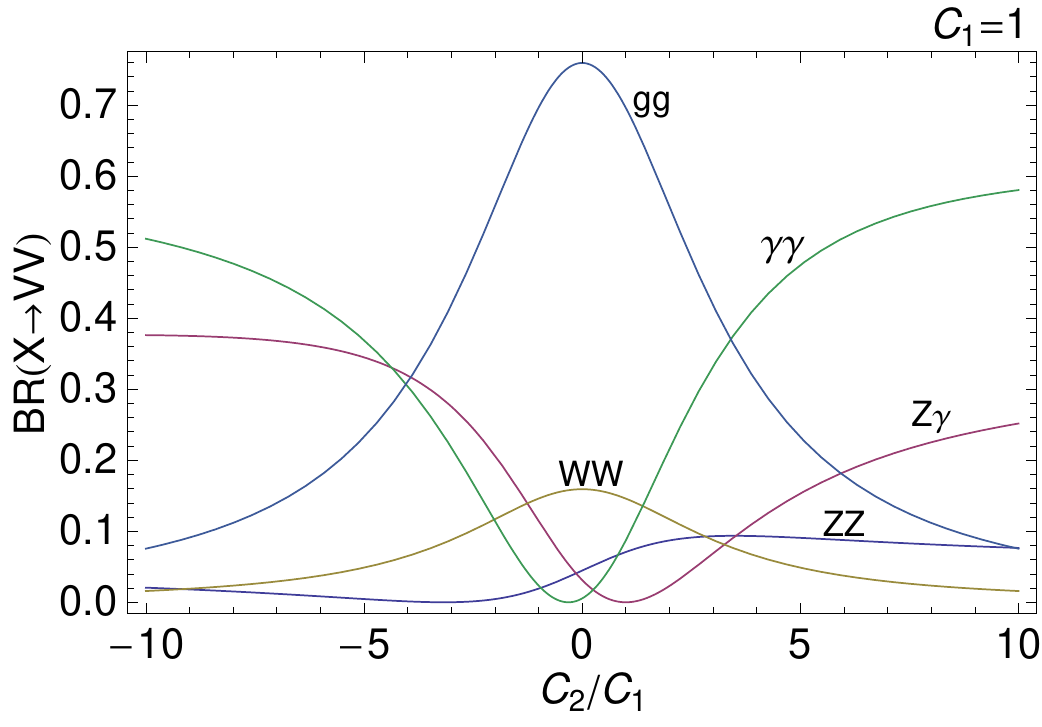} &
\includegraphics[width=3in]{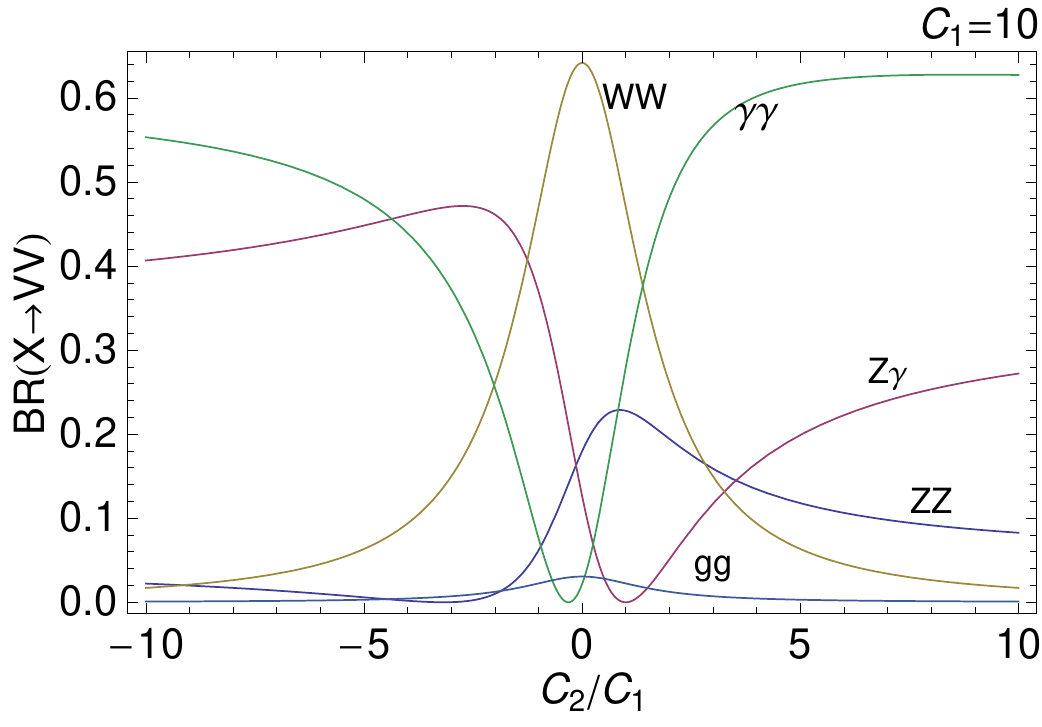} \\
\end{tabular}
\caption{$X$ decay branching fractions when $X$ is spin-0 as a function of $C_2 / C_1$.  Here, $m_X=500~\gev$ and
either $C_1 =1$ (left panel) or $ C_1 =10$ (right panel).  }
\label{fig:BRS}
\end{figure}

We can now calculate the branching fractions, again ignoring contributions from three-body decays, from Eqs.~(\ref{eq:swid})-(\ref{eq:pvwid}).
The branching fractions for spin-0 $X$ depend on two parameters, $C_1$ and $C_2$.
We illustrate this dependence in Fig.~\ref{fig:BRS} by plotting branching fractions for a certain choice of $C_1$ and as a function of $C_2 / C_1$.

\begin{figure}[h!]
\includegraphics[width=3in]{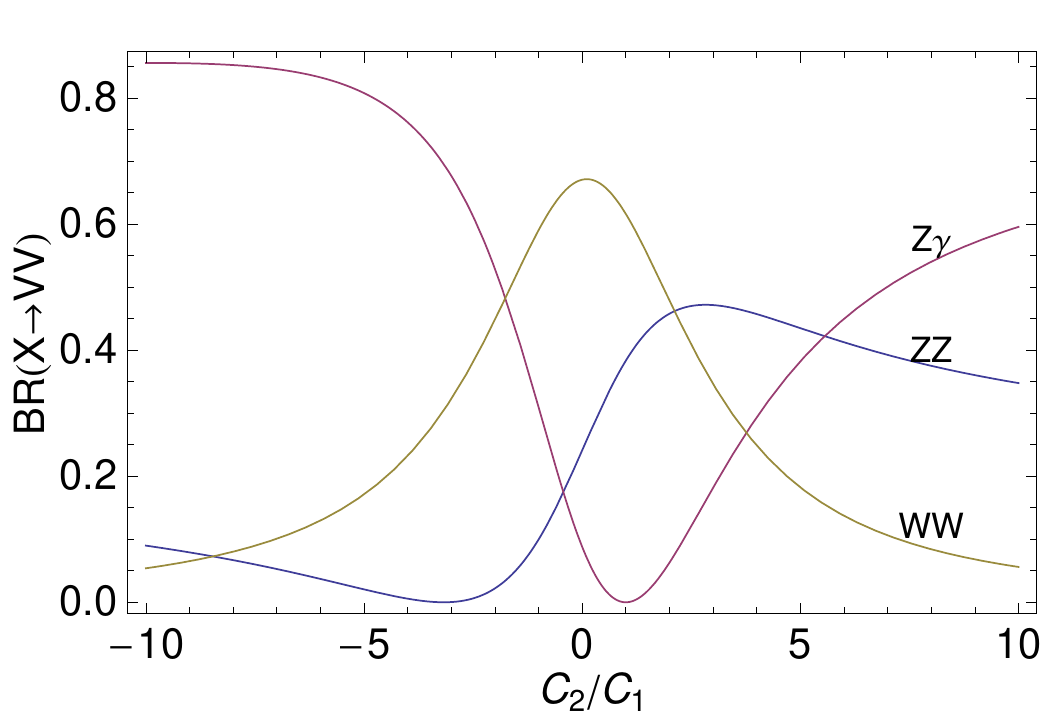}
\caption{$X$ decay branching fractions for when $X$ is spin-1 as a function of $C_2 / C_1$, for $m_X=500~\gev$.}
\label{fig:BRV}
\end{figure}

For the spin-1 $X$, the branching fraction to $gg$ (and $\gamma \gamma$) vanishes, and so the overall scale of
the electroweak couplings in relation to the gluon coupling drops out.  Thus the remaining branching fractions
only depend on $C_2 / C_1$, and are constant with respect to overall scale $\Lambda$.  These are shown in Fig.~\ref{fig:BRV}

Branching ratios in both cases are also dependent on $m_X$, but for $m_X \gtrsim 500~\gev$ the dependence becomes negligible.  Similarly,
differences between parity even and parity odd become negligible.

\subsection{Three-body decays}
In the previous section, we ignored three-body decay widths due to the usual assumption that the reduction in phase space
and extra coupling constant factors render these negligible.  There are cases, however, where this assumption cannot be made.
We show in Table~\ref{3BodyBR} the three-body branching ratios for all four spin/parity assignments at five different masses.
\begin{table}
\caption{Three-body branching fractions, as calculated in MadGraph5,  for $C_1 =10$, $C_2 =0$, and either
$\Lambda =50~\tev$ (spin-0) or $\Lambda =\sqrt{10}~\tev$ (spin-1).}
\label{3BodyBR}
\begin{tabular*}{0.75\textwidth}{@{\extracolsep{\fill}} l  c c c c c}
	\hline \hline
	 $m_X =$ & $500~\gev$ & $1000~\gev$ & $1500~\gev$ & $2000~\gev$ & $10~\tev$\\
	\hline
	$X_{s}$ & 0.078 & 0.154 & 0.207 & 0.24  & 0.436 \\
	$X_{ps}$ & 0.079 & 0.154 & 0.206 & 0.245 & 0.447 \\
	$X_{v}$ & 0.054 & 0.129 & 0.199 & 0.266  & 0.818 \\
	$X_{pv}$ & 0.055 & 0.129 & 0.199 & 0.267  & 0.816 \\
	\hline \hline
\end{tabular*}
\end{table}
We see that when $X$ is  spin-0, the three-body branching fractions are small, but in general they are not negligible.
More importantly, we see that if $X$ is  spin-1, the three-body decays strongly dominate for $m_X \gg m_{W,Z}$.
This behavior arises again from the Landau-Yang theorem.  For $m_X \gg m_{W,Z}$, the electroweak gauge bosons can be considered
approximately massless, and the two-body decay becomes highly suppressed.  We see this clearly in the vector and pseudovector
three-body branching fractions for $m_X = 10~\tev$.

It is thus important to note that searches for two-body resonances will be ineffective for very massive
fermiophobic spin-1 particles.  Furthermore, any search for a three-body resonance will be complicated by the fact that,
for electroweak decay products, full reconstruction using only leptons and photons is impossible.
The only three-body electroweak decay modes for the $X$ are $W^{+}W^{-}Z$ and $W^{+}W^{-} \gamma$; purely leptonic decays will
always produce missing momentum via neutrinos.  Thus for full reconstruction we must rely on the hadronic channels, (either
$X\rightarrow gq\bar{q}$ or hadronic decays of electroweak gauge bosons).  Since a vector $X$ must be produced with an
associated jet, this signal will be contained within events containing 4 or more jets.  Alternatively, a search can be performed in 
channels in which intermediate electroweak bosons are off-shell.

For $m_X \leq 2000~\gev,$ the branching fractions to electroweak gauge bosons dominate, and we may still be able to achieve discovery
in the two-body decay channels.  The remainder of this paper will focus on this possibility, and the possibility
of distinguishing spin if a resonance is indeed observed.

\section{Collider Simulation}\label{sec:ColAn}
As outlined above, we will generate event samples in which $X$ is produced through the $Xgg$ vertex and subsequently decays
through electroweak couplings.  We generate parton-level events in MadGraph5 \cite{Alwall:2011uj}, shower and hadronize the events
in Pythia \cite{Sjostrand:2006za}, and perform detector simulation using PGS4 \cite{PGS4} using the ATLAS detector card.
To generate the model files for MadGraph, we have used the Mathematica package FeynRules \cite{FeynRules}, to which we
feed the effective interaction operators, Eq.~(\ref{eq:OPs})-(\ref{eq:OPpv}), as input.
We generated events for collider energies of $\sqrt{s}=8~\tev$ and $\sqrt{s}=14~\tev$.

For spin-0 $X$, we generate both $p p \rightarrow X$ and $pp \rightarrow X j$ at the parton level, in order to include all possible production channels.  These two processes are matched at the Pythia level using the MLM algorithm in order to avoid over counting between matrix element generated and ISR generated jet-containing events.  For the spin-1 cases, events are generated solely from the 1-jet matrix element (which in this case is tree-level).

\subsection{Cuts}
The following reconstruction cuts will be applied for the three channels we study.  These cuts are applied at the detector level.

\begin{itemize}
\item{$X \rightarrow ZZ$

We require events to contain 4 charged leptons ($e^\pm$, $\mu^\pm$) with $\eta \leq 2.5$.
These must consist of two pairs of same-flavor opposite-sign leptons which each
have invariant masses in the range $80~\gev \leq m_{l^+ l^-} \leq 100~\gev$.
The total invariant mass must lie within $10 \% $ of $m_X$.  }

\item{$X \rightarrow Z\gamma$

We require events to contain a photon as well as 2 same-flavor opposite-sign leptons ($\eta \leq 2.5$) whose
invariant mass lies in the range $80~\gev \leq m_{l^+ l^-} \leq 100~\gev$.
The two leptons and photon then must have a total invariant mass which lies within $10 \%$ of $m_X$.  }

\item{$X \rightarrow \gamma \gamma$

We require events to contain 2 photons ($\eta \leq 2.5$) which reconstruct to invariant mass within $10 \%$ of $m_X$.  }
\end{itemize}

In addition to these cuts, a number of standard isolation cuts are applied at the MadGraph level, including requiring $\Delta R > 0.4$ for any pair of particles.
At this point we do not impose any cuts on the jet structure of the events.

\subsection{Background}
We generate background events using the same MadGraph5 simulation chain.
For the vast majority of events where a lepton pair has an invariant mass between $80-100~\gev$, the leptons
arise from the decay of a $Z$ boson.  Other sources of lepton pairs satisfying the cuts (including lepton
misidentification) are insignificant and can be ignored.
For the $ZZ$ and $Z\gamma$ background channels, we use 0-jet and 1-jet matrix elements,
which are matched using the MLM algorithm.  For the $\gamma \gamma$ channel, we use only the 0-jet matrix element.
Cross sections for these channels, after imposing the above mentioned cuts in each $X$-mass window, are given in Table~\ref{BGxsec} at LHC
energies of $8~\tev$ and $14~\tev$ (see also~\cite{Baur:1988cq,Baur:1989cm}).

\begin{table}[h!]
\caption{Inclusive cross sections (in fb) for standard model background in the $ZZ,$ $Z \gamma,$ and $\gamma \gamma$ channels,
after reconstruction cuts in each total invariant mass window.}
\label{BGxsec}
\begin{tabular}{|c|l|c|c|c|c|}
	\hline
\multicolumn{2}{|c|}{\,}  & \multicolumn{4}{c|}{$m_X$} \\
\cline{3-6}
	\multicolumn{2}{|c|}{\,}  & 500 GeV & 1000 GeV & 1500 GeV & 2000 GeV \\
	\hline
%	8 TeV\\
	& $\sigma_{BG}(pp \rightarrow ZZ \rightarrow l^+ l^- l^+ l^-)$ & $2.09\times 10^{-1} $ & $7.35 \times 10^{-3}$
& $1.87\times 10^{-4}$ & $2.16\times 10^{-5}$ \\
	8 TeV & $\sigma_{BG}(pp \rightarrow Z \gamma \rightarrow l^+ l^- \gamma )$ & 3.93 & $2.98\times 10^{-1}$ & $2.55 \times 10^{-2}$ & $2.93\times 10^{-3}$\\
	& $\sigma_{BG}(pp \rightarrow \gamma \gamma)$ & $1.80\times 10^{1}$ & $1.42$ & $2.08\times 10^{-1} $ & $4.66\times 10^{-2}$ \\
	\hline
%	14 TeV \\
	& $\sigma_{BG}(pp \rightarrow ZZ \rightarrow l^+ l^- l^+ l^-)$ & $4.41\times 10^{-1}$ & $2.45\times 10^{-2}$
& $1.90 \times 10^{-3}$ & $1.61 \times 10^{-4}$ \\
	14 TeV & $\sigma_{BG}(pp \rightarrow Z \gamma \rightarrow l^+ l^- \gamma )$ & 8.42 & $4.16\times 10^{-1}$ & $1.09\times 10^{-1}$ & $2.72\times 10^{-2}$\\
	& $\sigma_{BG}(pp \rightarrow \gamma \gamma)$ & $3.65\times 10^{1}$ & $2.79$ & $8.73\times 10^{-1}$ & $3.27\times 10^{-1}$ \\
	\hline
\end{tabular}
\end{table}

\section{Collider Reach}\label{sec:reach}

Since the narrow width approximation holds, the number of signal events in a given channel and at a given luminosity ${\cal L}$ depends on the
coupling constants $C_1$ and $C_2$ only though the final state branching fractions.  If we denote by $\sigma_{\rm prod}$ the cross-section
for the production process $pp \rightarrow X+({\rm 0\,and/or\,1-jet})$, and by $\epsilon$ the fraction of such events which pass the analysis
cuts in any channel, then the number of signal events  is given by
\begin{align}
N_{sig} &=\sigma_{\rm prod} \times \epsilon \times BR(X\rightarrow VV)\times {\cal L} \nonumber \\
&\propto\Lambda^{-n} \times BR(X\rightarrow VV)\times {\cal L} \label{eqn:Nevents}
\end{align}
where $n=2$ for spin-0, and $n=4$ for spin-1.
For a discovery, we will require a $5\sigma$ (Gaussian equivalent) excess of signal events over background events satisfying the
same cuts, and we require at least 5 signal events.
With Eq.~(\ref{eqn:Nevents}) in mind, we define the LHC reach for any integrated luminosity in terms of the quantity
\bea
R_{VV} &\equiv& \Lambda / [ BR(X\rightarrow VV) ] ^{1/n} .
\eea

We find the reach for $10~\ifb$ and $30~\ifb$ at center of mass energy of $\sqrt{s}=8~\tev$, and $100~\ifb$ at $\sqrt{s}=14~\tev$.
For the scalar and vector cases, the LHC reach
is plotted against $m_X$ in Fig.~\ref{fig:reach}.
We plot the inclusive $X \rightarrow ZZ$ channel (both 0 and 1-jet events) and the
inclusive $X \rightarrow Z \gamma$ channels separately.
The reach plots for the
pseudoscalar and pseudovector case are essentially identical to the scalar and vector cases, respectively.
\begin{figure}[h!]
\begin{tabular}{cc}
\includegraphics[width=3in]{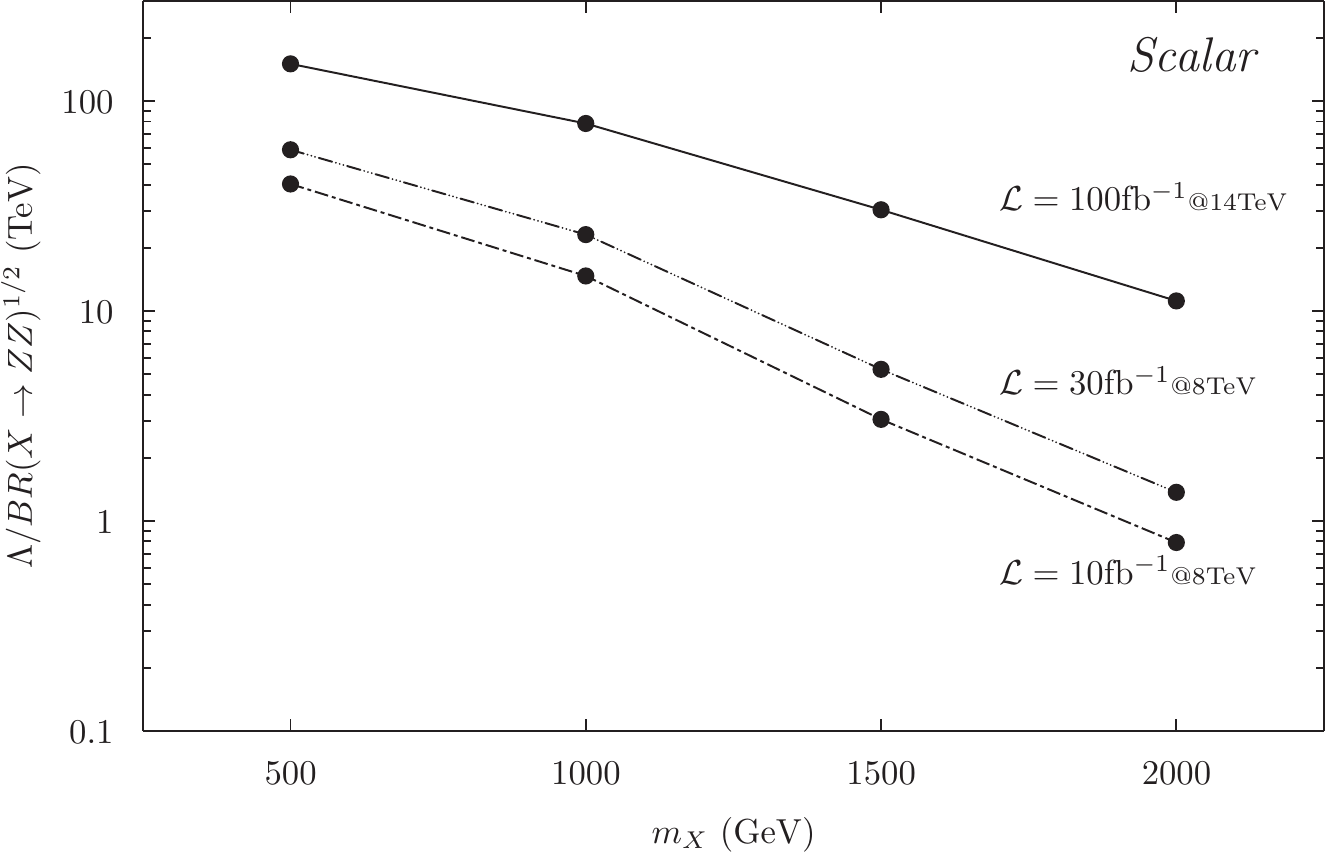} &
$~~$ \includegraphics[width=3in]{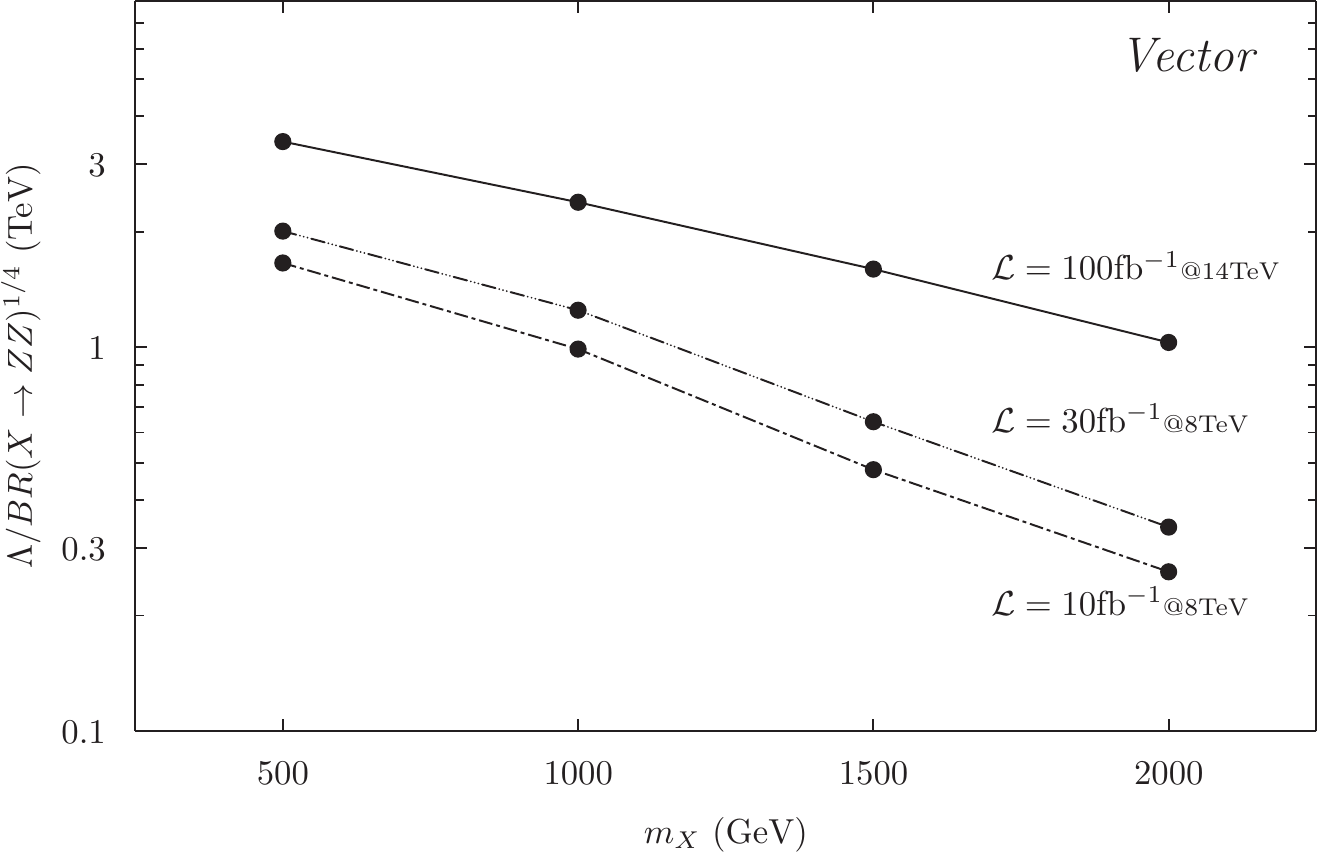} \\
\includegraphics[width=3in]{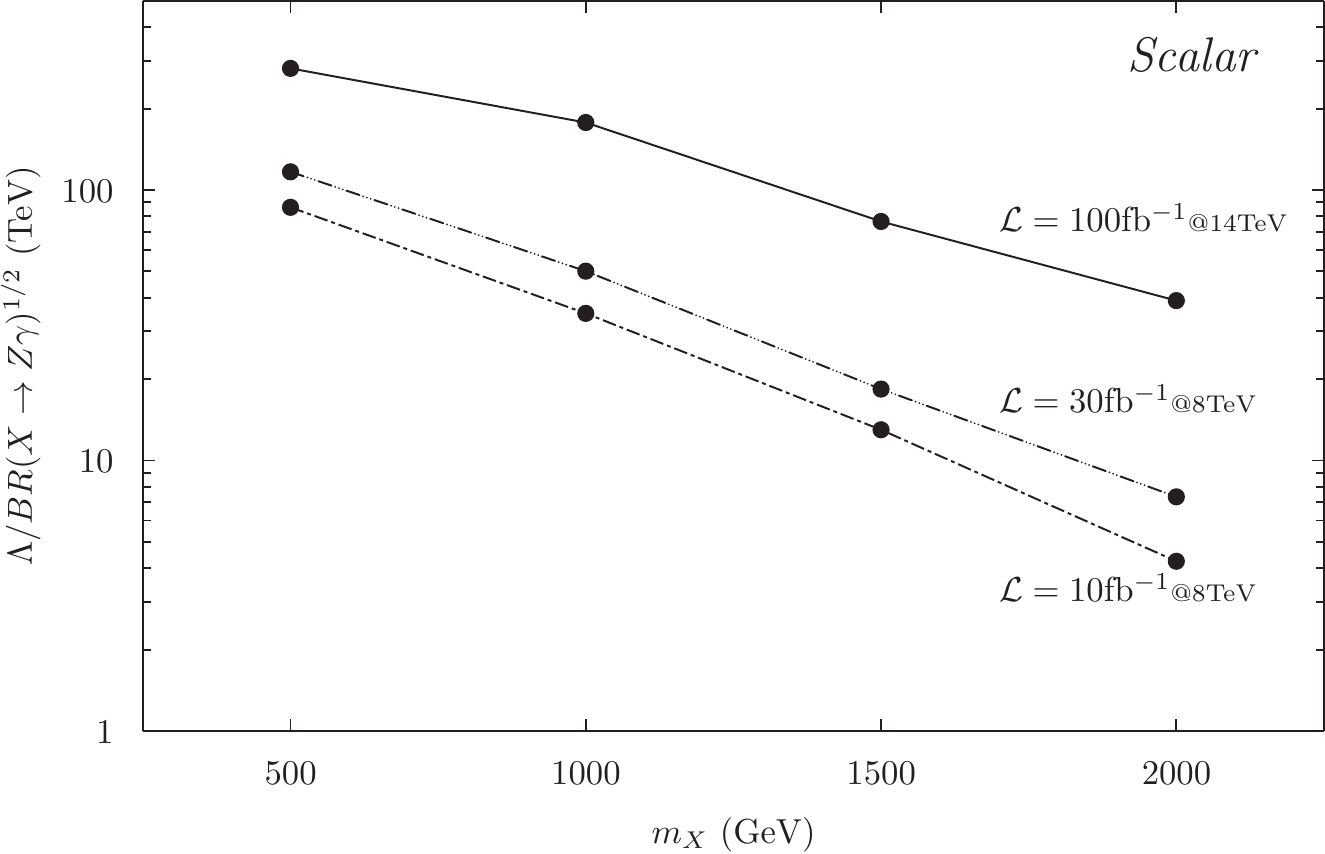} &
$~~$\includegraphics[width=3in]{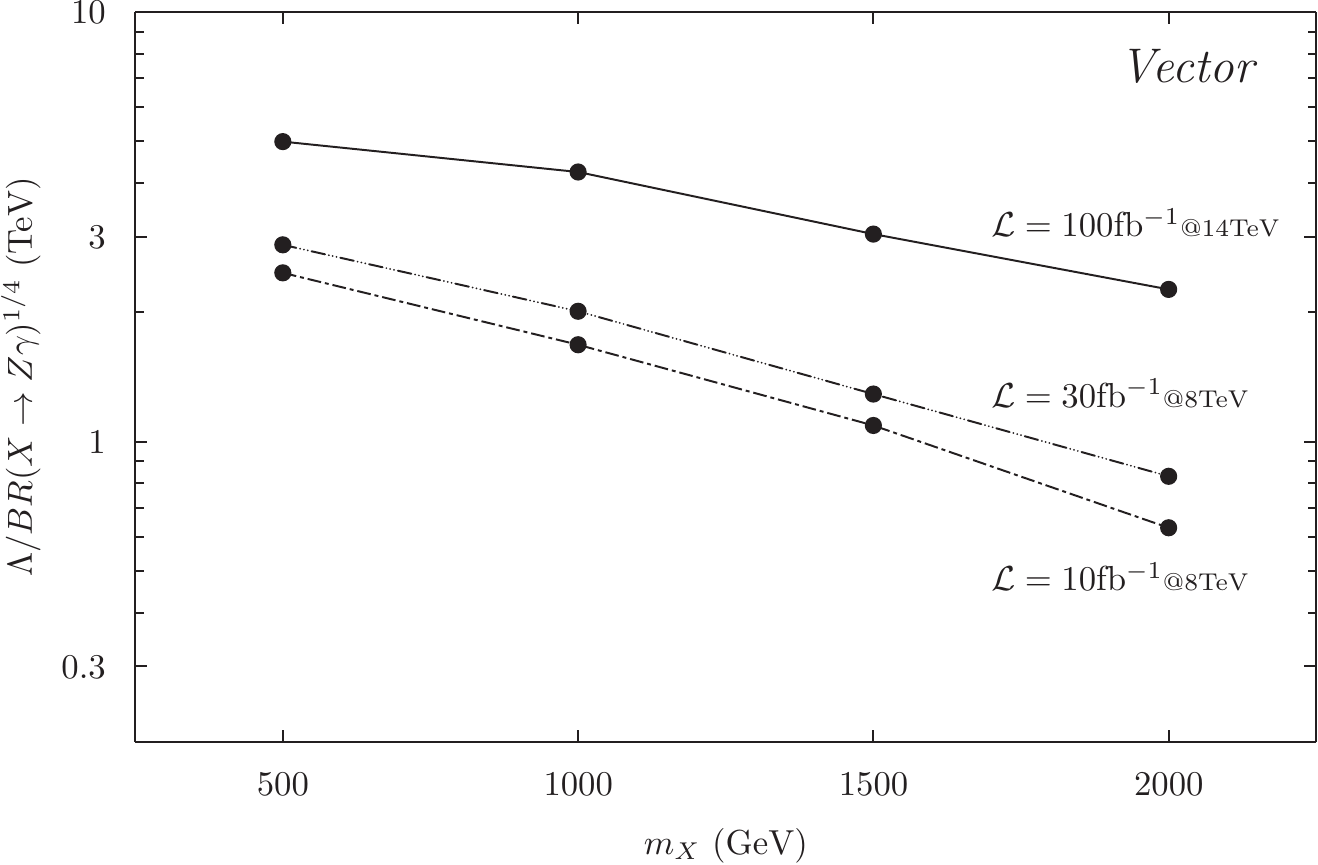}
\end{tabular}
\caption{Collider reach for both $ZZ$ (top) and $Z \gamma$ (bottom) channels.  Spin-0 $X$ is shown at left, spin-1 $X$ at right.  The reach plots for the
parity odd bosons are essentially identical to their parity even counterparts.}
\label{fig:reach}
\end{figure}
We see that in both channels, we can probe higher scales in the spin-0 case, which is expected from the additional $\Lambda^2$
 suppression coming from the dimension 6 operators coupling a spin-1 fermiophobic boson to the standard model.
 We also note that the $Z\gamma$ channel is generally a better probe despite the higher
 backgrounds.  This is mainly due to the small $Z \rightarrow l^+ l^-$ branching fraction and the reconstruction cuts present in the $ZZ$ channel.

In addition, we show in Fig.~\ref{fig:reachAA} the reach for the $\gamma \gamma$ channel, present only for spin-0 $X$.  Current
LHC data can probe well into the hundreds-of-TeV range.

\begin{figure}[h!]
\includegraphics[width=3in]{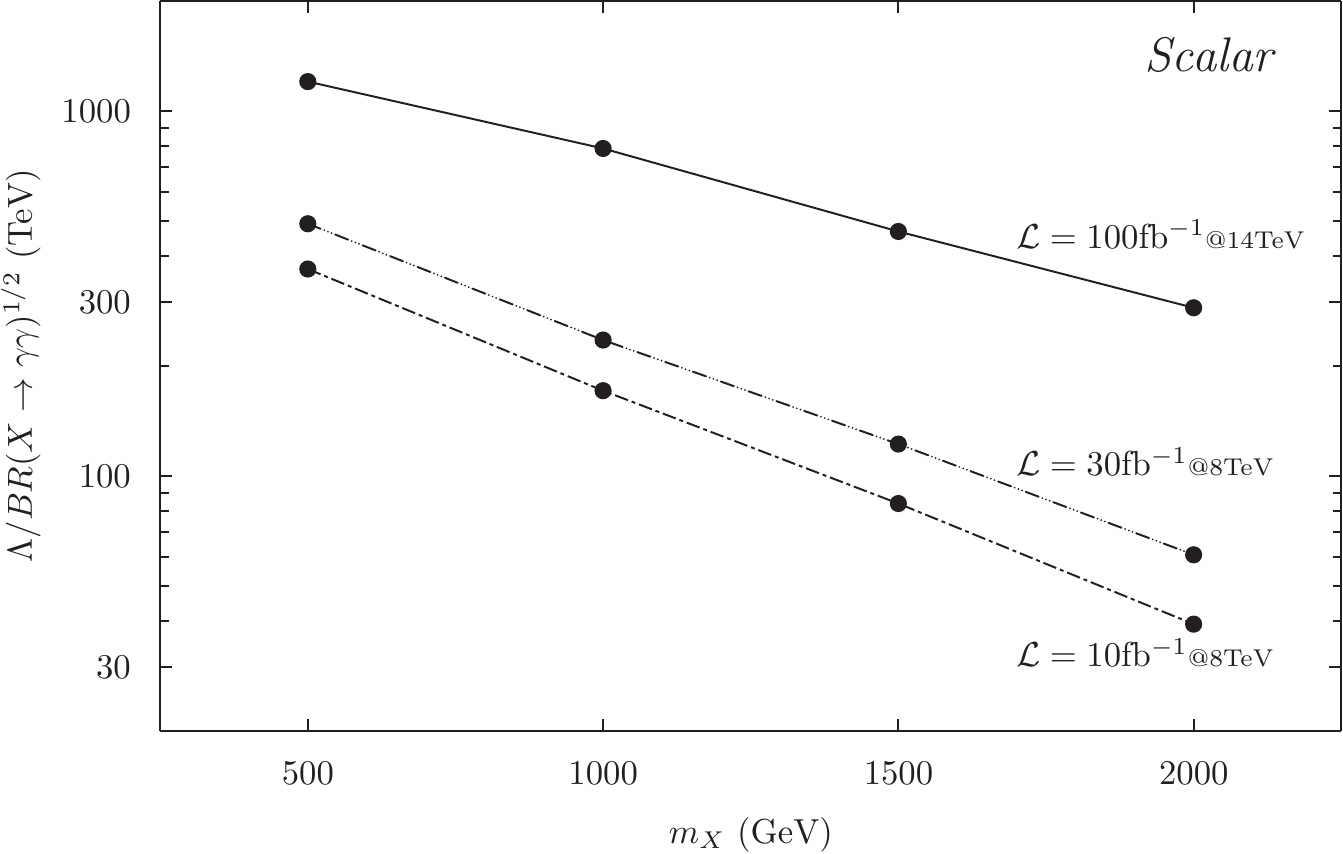}
\caption{Collider reach for both $\gamma \gamma$ channel, which is present only for spin-0 $X$.}
\label{fig:reachAA}
\end{figure}

Note that we have not accounted for NLO QCD corrections to the $X$ production cross section.  The
associated $K$-factors are usually greater than one, indicating that a correct treatment of NLO effects
would likely increase the LHC reach.

\subsection{Spin Determination}
It is clear from the reach plots of the previous section that the simplest way to distinguish a spin-0 resonance from spin-1
is to look to the $\gamma \gamma$ channel. If there is
an observation in the $\gamma \gamma$ channel, with or without a corresponding observation in $ZZ$ or $Z \gamma$, the Landau-Yang
theorem implies that the decaying
particle is not spin-1.

There are cases, however, where a spin-0 $X$ will not decay to two photons.  We see from Fig.~\ref{fig:BRS} and Eq.~(\ref{eq:sWidAA}) that the branching
ratio $BR(X\rightarrow \gamma \gamma)$ drops to zero for \mbox{$C_2 /C_1=- \tan^2\theta_w$} (an explicit model where this occurs is exhibited in
the Appendix).
The electroweak decay channels for the spin-0 and spin-1 $X$ are in this case identical.  We must in these cases resort to an alternative method of determining spin.

Another way of distinguishing between spin-0 and spin-1 resonances is from the presence of extra jets; the 0-jet process
$pp \rightarrow X \rightarrow ZZ,Z\gamma$ is
possible if $X$ is spin-0, but not if it is spin-1.  As we saw, we can define the LHC reach for $X$ (spin-0) in terms of the quantity
$R_{VV} = \Lambda / BR(X \rightarrow VV)^{1/2}$.  We thus see that a search in the $ZZ$ or $Z\gamma$ channel with zero extra jets will be more
promising than a search in the $\gamma \gamma$ channel if
\bea
{R_{ZZ,Z\gamma}(\mbox{0-jet}) \over R_{\gamma \gamma}}\left[{BR(X\rightarrow ZZ,Z\gamma) \over BR(X\rightarrow \gamma \gamma) }
\right]^{1\over 2} &>& 1.
\eea
The ratios $BR(X\rightarrow ZZ)/BR(X\rightarrow \gamma \gamma)$ and $BR(X\rightarrow Z\gamma)/BR(X\rightarrow \gamma \gamma)$
depend only on the quantities $m_X$ and  $C_2 / C_1$.
Having determined $R_{ZZ}(m_X)$ and $R_{Z\gamma}(m_X)$ for the 0-jet sample from the generated signal and background events, one can
determine the range of $C_2 / C_1$ (for any $m_X$) over which the $ZZ$ or $Z\gamma$ 0-jet search provides better prospects for
discovery than the $\gamma \gamma$ channel.  These ranges are plotted in Fig.~\ref{fig:C2C1noAA}; note that these ranges would
not change significantly after the LHC energy and luminosity upgrade.

\begin{figure}[h!]
\begin{tabular}{cc}
\includegraphics[width=3in]{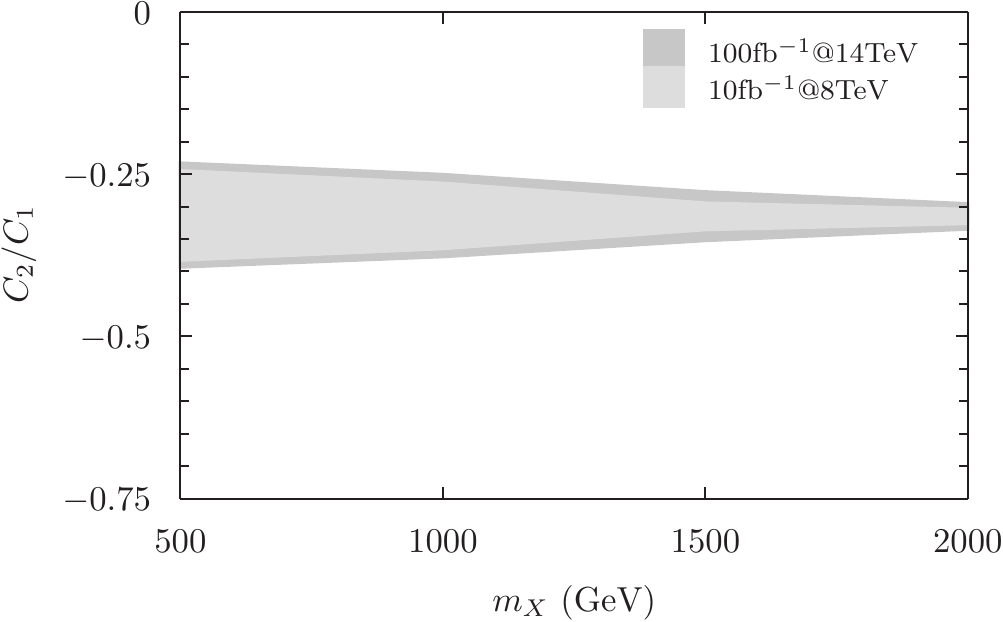}&
\includegraphics[width=3in]{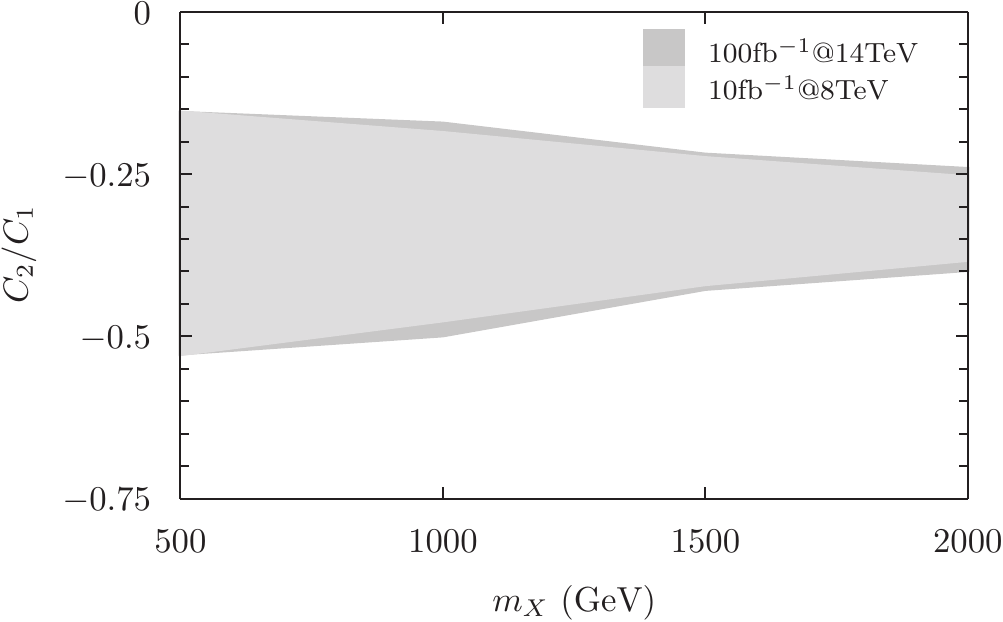}
\end{tabular}
\caption{Range of $C_{2} / C_{1}$ for which a $ZZ$-channel search (left) or a $Z\gamma$-channel (right) search is more promising than a diphoton search.}
\label{fig:C2C1noAA}
\end{figure}

Thus far, our analysis has focused on strategies for determining if $X$ is not spin-1.
One might ask the converse question: does the appearance of an excess in $ZZ$ and/or $Z\gamma$ events with one
extra jet, unaccompanied by an excess in 0-jet events or $\gamma \gamma$ events, necessarily
imply the $X$ resonance is spin-1?  Essentially, this amounts to the question of whether or not a
spin-0 resonance can produce an excess in 1-jet events, without also producing a statistically
significant excess in 0-jet events or $\gamma \gamma$ events.

We can determine this by comparing the LHC reach for 0-jet events to the reach for 1-jet events in the $ZZ$ or
$Z\gamma$ events, assuming $X$ is spin-0.  If the reaches are comparable, then a spin-0 $X$ coupling to the
standard model in such a way as to produce
a $5\sigma$ excess in $ZZ$ or $Z\gamma$ events with 1 extra spectator jet would, for similar luminosity, also produce an
excess in events with no spectator jets.  An observed $5\sigma$ excess in 1-jet events without some excess in
0-jet events would imply that $X$ is not spin-0.

In Fig.~\ref{fig:R0R1} we plot $R_{VV}(\mbox{0-jet})/R_{VV}(\mbox{1-jet})$ as a function of $m_X$ for the $ZZ$ and $Z\gamma$ channels
(assuming $X$ is spin-0).  From this plot, we see that for much of the range, an observation of a resonance
with one spectator jet, but an absence of a excess with zero spectator jets, is sufficient to demonstrate that the
resonance is not spin-0.  Furthermore, we see that this conclusion is largely independent of the collider energy, luminosity, and decay channel studied.
\begin{figure}[h!]
\begin{tabular}{cc}
\includegraphics[width=3in]{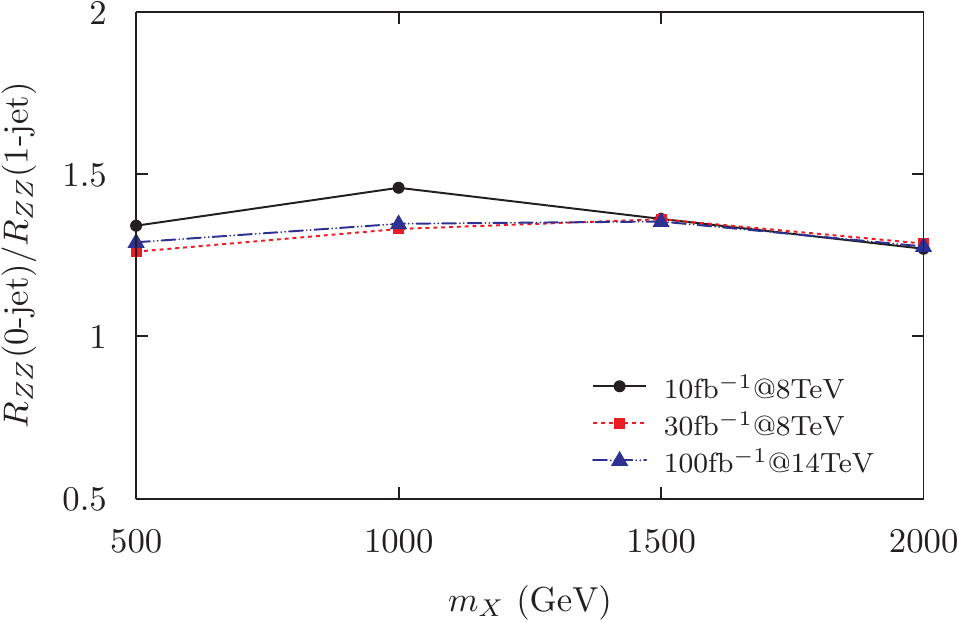}&
\includegraphics[width=3in]{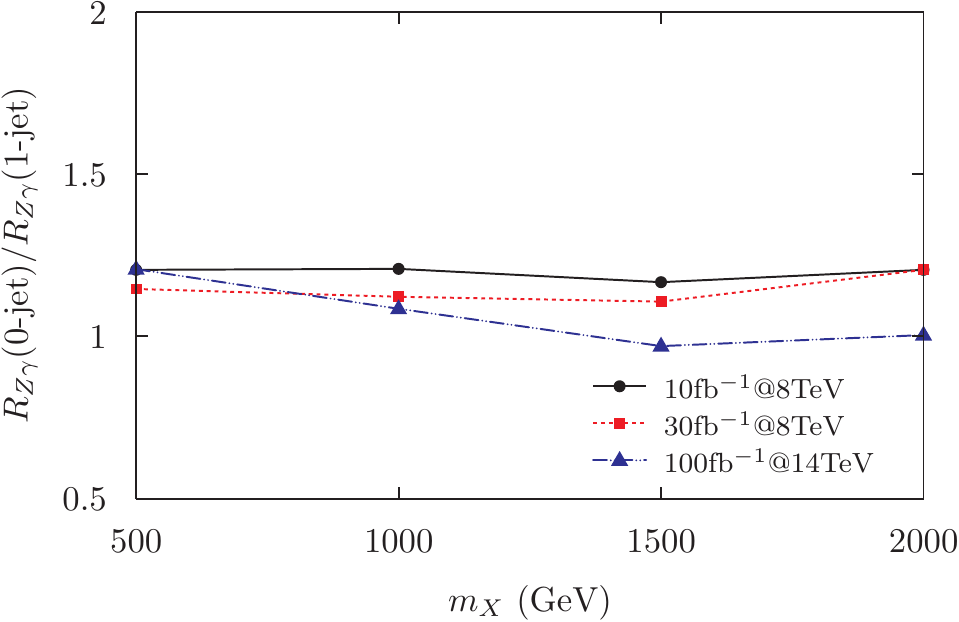}
\end{tabular}
\caption{Ratio of reaches $R_{VV}(\mbox{0-jet})/R_{VV}(\mbox{1-jet})$ for $X \rightarrow ZZ$ (left) and $X \rightarrow Z\gamma$
(right) in the case of spin-0 $X$.  We see that if the resonance is spin-0, we expect a similar number of 0-jet and 1-jet events.}
\label{fig:R0R1}
\end{figure}
We also show (Fig.~\ref{fig:R0R1vect}) $R_{VV}(\mbox{0-jet})/R_{VV}(\mbox{1-jet})$ for the case of a spin-1 $X$ (in this case, the
number of signal events is proportional to $R^{-4}$).  As expected,
we see that an observation in the 0-jet channel without a corresponding observation in the 1-jet
channel is sufficient to demonstrate that the resonance is not spin-1.
\begin{figure}[h!]
\begin{tabular}{cc}
\includegraphics[width=3in]{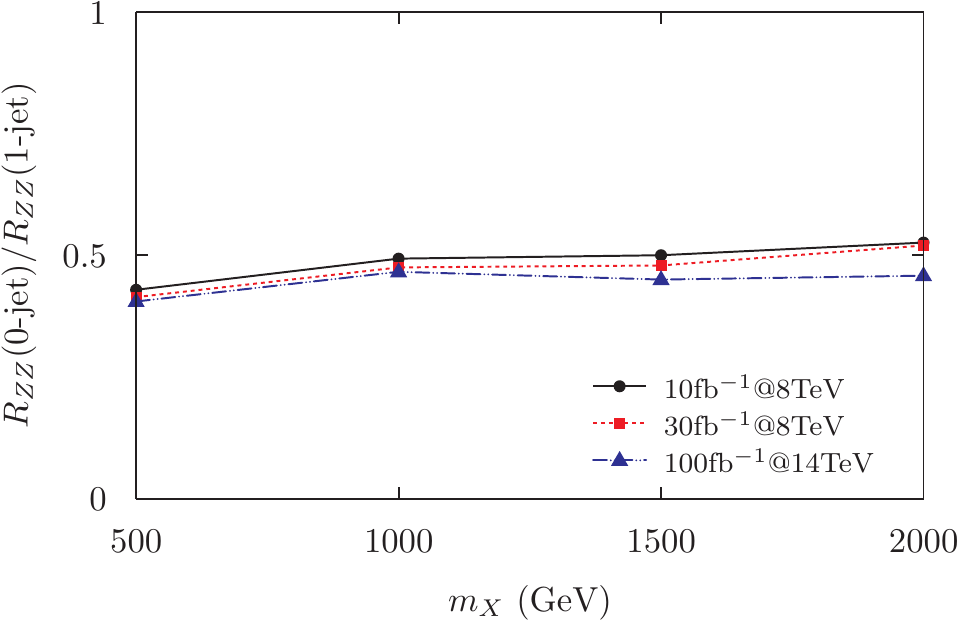}&
\includegraphics[width=3in]{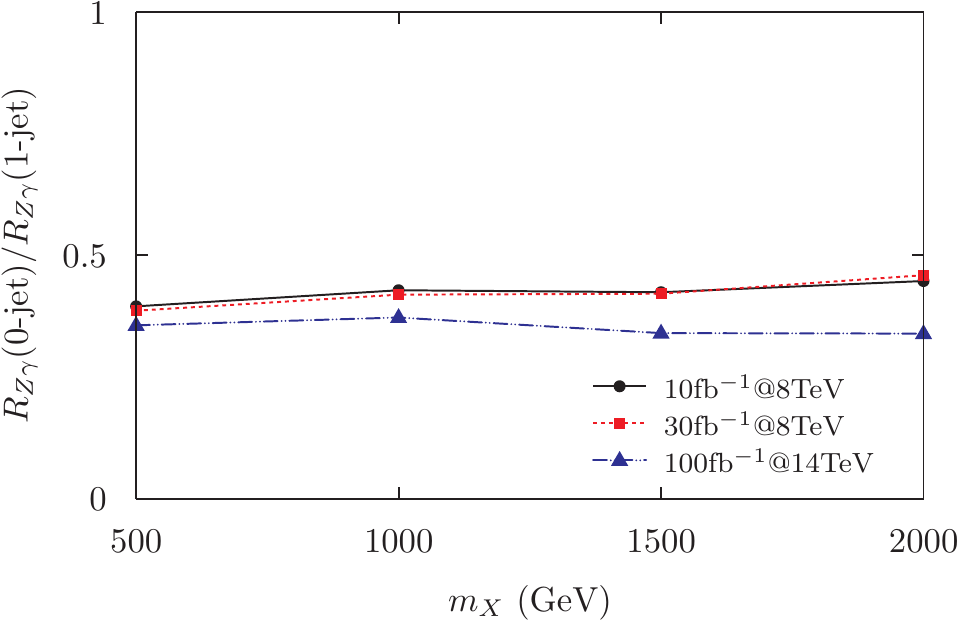}
\end{tabular}
\caption{Ratio of reaches $R_{VV}(\mbox{0-jet})/R_{VV}(\mbox{1-jet})$ for $X \rightarrow ZZ$ (left) and $X \rightarrow Z\gamma$
(right) in the case of spin-1 $X$.  We see that if the resonance is spin-1, we expect to many 1-jet events and very few 0-jet events.}
\label{fig:R0R1vect}
\end{figure}

\section{Conclusions}

The existence of exotic fermiophobic bosons is well-motivated theoretically.  These types of bosons
could reveal themselves as resonances which decay to two or more SM bosons at the LHC.  We have established
the efficacy of the LHC in probing models of this type, and demonstrated the viability of two methods for
determining spin.  The existence of a diphoton decay channel establishes that the resonance is not spin-1, as is well-known.
When a boson does not decay to two photons, spin can be determined through an inspection of the jet structure
of the signal events.  Over much of the parameter space, a discovery and a spin determination can be achieved
in as few as 5 events, thus requiring far fewer statistics than alternative methods of spin determination (\eg~analyses
of angular distributions of decay products).

The LHC has already discovered a new boson, and the existence of diphoton decays shows definitively that
this new particle is not spin-1.  We have shown that the LHC has very good prospects for discovering fermiophobic bosons in the near future,
even if they are relatively heavy, or are only coupled to the standard model by higher dimension effective operators
which are heavily suppressed.  For example, a $100~\ifb$ run of the LHC at $14~\tev$ could find $5\sigma$ evidence for
a scalar with $m_X \sim 2~\tev$, coupled to the standard model by effective operators suppressed by a mass scale
$\sim 300~\tev$.
The reason for this large reach is that a fermiophobic boson can be produced from
gluon couplings, but observed through electroweak decays, a channel which is ideal for detection at the LHC.
In particular, the diphoton channel is clearly the most promising, as the signal is very clean and the background
is very small.

But one typically expects an electrically
neutral boson to have a very small branching fraction for decay to $\gamma \gamma$.  This is the case for the $125~\gev$
boson discovered at the LHC.
Fermiophobic scalars are a major contrary example; the branching fraction for diphoton decay can easily be
${\cal O}(1)$.  One can see why such large $X \rightarrow \gamma \gamma$ branching fractions are allowed by 
considering a high-energy theory where the fermiophobic boson $X$ couples to standard model gauge bosons 
only through loops of heavy fermions and scalars which are charged under standard model gauge groups.  In this 
scenario, since all decays arise from one-loop diagrams, the $X \rightarrow \gamma \gamma$  branching fraction 
can be comparable to that of other channels.  This behavior is markedly different from other scenarios in which 
some decays to standard model particles occur at tree-level.
Fermiophobic bosons are thus one of the most interesting prospects for future discovery at an
upgraded LHC.

Although the methods we have discussed here are applicable for spin determination,
it is clear from our results that parity information is
generally not encoded in jet number distributions.  Alternative methods must be used to determine the parity of the
couplings, which would give clues to the nature of the underlying physics of the new resonance.

\vskip .2in
\textbf{Acknowledgments}

We are grateful to B.~Thomas and J.~Alwall for useful discussions.
The work of J.~K. and D.~Y. is supported
in part by Department of Energy grant DE-FG02-04ER41291.
The work of A.R. is supported in part by NSF grant PHY-0970173.

\appendix

\section{Vanishing $BR(X \rightarrow \gamma \gamma)$}

The condition that  $C_2=-C_1 \tan^2\theta_w,$ which is necessary for a vanishing branching ratio to two photons, can arise naturally as follows.  Let us assume that the $X$ couples to electroweak
gauge bosons through triangle diagrams involving extra heavy fermions.  In this model there are two extra vector-like fermion multiplets.
One is an SU(2) doublet,
\[
\left( \begin{array}{c}
a \\
b
\end{array} \right),  \text{ with } Y=-1,
\]
which couples to $X$ through a Yukawa interaction with coupling constant $\lambda_1$. The other is an SU(2) singlet, $c$,
with $Y=-2$ and which couples to $X$ through a Yukawa interaction with coupling constant $\lambda_2$.
The electric charge ($Q=T^3 + Y/2$) of $b$ and $c$ is $-e$ while the charge of the $a$ is zero.

Let us first consider the coupling to two photons.  Only $b$ and $c$ will contribute to the loop, so the amplitude is proportional to
\[
\raisebox{-8mm}{\includegraphics[width=1.9in]{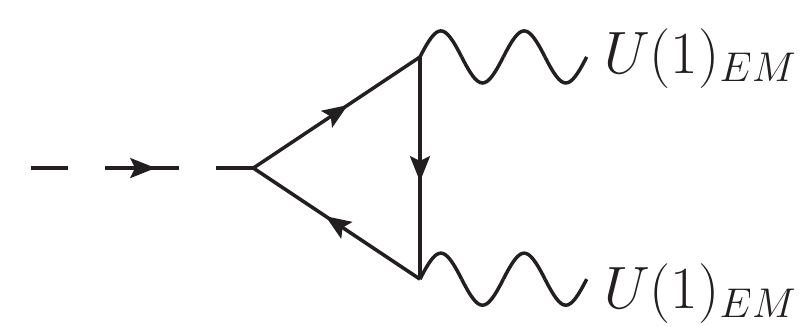} }
\begin{aligned}
\propto \lambda_1 (-e)^2 + \lambda_2 (-e)^2
\end{aligned}
\]
We see that for the $X\gamma\gamma $ coupling to vanish, we must have $\lambda_1 = - \lambda_2$.  Now, we look to the couplings to the
SU(2) and U(1)$_Y$ gauge bosons.  For SU(2), only the doublet runs in the loop, and we have
\[
\raisebox{-4mm}{\includegraphics[width=1.9in]{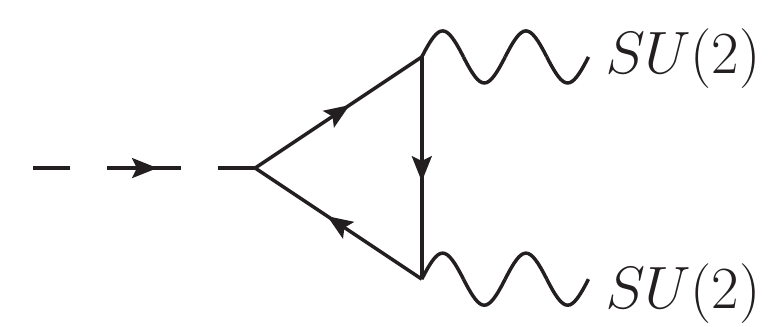} }
\begin{aligned}
&\propto \lambda_1 g^2 \Tr(t_i t_j) \\
&\propto \tfrac{1}{2}\lambda_1 g^2
\end{aligned}
\]
We can identify this factor as our coefficient $C_1$.  For U(1)$_Y$, all three particles flow in the loop, and we find
\[
\raisebox{-2mm}{\includegraphics[width=1.9in]{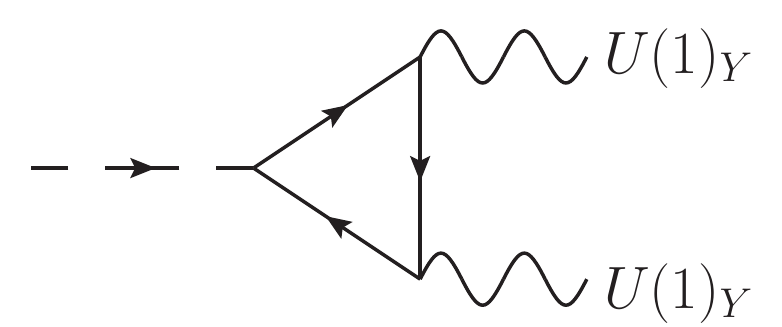} }
\begin{aligned}
&\propto \lambda_1 g^{\prime 2}(-\tfrac{1}{2})^2 +  \lambda_1 g^{\prime 2}(-\tfrac{1}{2})^2 +  \lambda_2 g^{\prime 2}(-1)^2 \\
&= \tfrac{1}{2}\lambda_1 g^{\prime 2} +  \lambda_2 g^{\prime 2} \\
&= -\tfrac{1}{2}\lambda_1  g^{\prime 2},
\end{aligned}
\]
where in the last line we replaced $\lambda_2=-\lambda_1$.
This coefficient can be identified with $C_2$.  Since $g^{\prime} = g \tan \theta_w,$ we see that $C_2 = -C_1 \tan^2 \theta_w$ as needed.

The above argument is strictly only valid if the particles coupling to the X --- $a$, $b$, and $c$ --- have identical masses.
If this condition is relaxed, then the diphoton coupling will vanish only for $\lambda_1 \approx - \lambda_2$.

\end{document}